\documentclass[aps,amsmath,superscriptaddress,preprint]{revtex4}
\usepackage{graphicx}
\graphicspath{{./final/}}

\begin{document}

\title{Long range proximity effect in $\textrm{L\ensuremath{a_{2/3}}C\ensuremath{a_{1/3}}Mn\ensuremath{O_{3}}}$
/(100)$\textrm{YB\ensuremath{a_{2}}C\ensuremath{u_{3}O_{7-\delta}}}$
ferromagnet/superconductor bilayers: Evidence for induced triplet
superconductivity in the ferromagnet.}

\author{Yoav Kalcheim}
\affiliation{Racah Institute of Physics and the Hebrew University
Center for Nanoscience and Nanotechnology, The Hebrew University of
Jerusalem, Jerusalem 91904, Israel }

\author{Tal Kirzhner}
\author{Gad Koren}
\affiliation{Department of Physics, Technion-Israel Institute
of Technology, Haifa 32000, Israel }

\author{Oded Millo}
\affiliation{Racah Institute of Physics and the Hebrew University
Center for Nanoscience and Nanotechnology, The Hebrew University of
Jerusalem, Jerusalem 91904, Israel }

\begin{abstract}
Scanning tunneling spectroscopy measurements conducted on epitaxially
grown bilayers of half-metallic ferromagnetic $\textrm{L\ensuremath{a_{2/3}}C\ensuremath{a_{1/3}}Mn\ensuremath{O_{3}}}$
(LCMO) on superconducting (SC) (100)$\textrm{YB\ensuremath{a_{2}}C\ensuremath{u_{3}O_{7-\delta}}}$
(YBCO) reveal long-range penetration of superconducting order into
the LCMO. This anomalous proximity effect manifests itself in the
tunneling spectra measured on the LCMO layer as gaps and zero bias
conductance peaks. Remarkably, these proximity-induced spectral features
were observed for bilayers with LCMO thickness of up to 30 nm, an
order of magnitude larger than the expected ferromagnetic coherence
length in LCMO. We argue that this long-range proximity effect can
be accounted for by the formation of spin triplet pairing at the LCMO
side of the bilayer due to magnetic inhomogeneity at the interface
or at domain walls. Possible symmetries of the induced order parameter
are discussed.
\end{abstract}
\maketitle

\section*{Introduction}

Significant research has been dedicated lately to the characterization
of half metallic ferromagnets (HMF) motivated by their application
in spintronics as sources of spin polarized current. The proximity
effect in HMF \textendash{} superconductor (SC) junctions has also
drawn much attention recently due to its possible application in magnetization
controlled Josephson junctions\cite{Keizer2006} and more generally,
as means for investigating the interplay between the competing orders
of ferromagnetism (F) and SC. The underlying mechanism of the proximity
effect (PE) in a normal metal (N) in good electrical contact with
a SC is the Andreev reflection (AR) whereby a hole-like quasiparticle
from the N side impinging on the interface is retro-reflected as an
electron-like quasiparticle with opposite spin, destroying a Cooper
pair on the SC side. By this, superconductivity is reduced in the
SC side of the interface and pairing correlations and thus a SC order
parameter (OP) are induced in N. At a finite temperature, T, superconducting
correlations will be observed in N up to a distance $\xi_{N}=\sqrt{\hbar D/k_{B}T}$
from the N-SC interface, where D is the diffusion coefficient. Beyond
this so called 'normal coherence length', typically of the order of
ten of nm, the hole-like and electron-like quasiparticles start losing
phase coherence. The AR process, or the PE, are naturally expected
to be drastically suppressed at F-SC interfaces due to the exchange
interaction in F. Indeed, by applying the methodology originally introduced
by Fulde, Ferrel, Larkin and Ovchinikov (FFLO)\cite{FuldeFerrel,LarkinOvchinokov}
for treating magnetic superconductors to the problem of F-SC junctions,
it was shown\cite{DemlerBeasleyPEinF1997,BuzdinPEinSC-F2005} that
the exchange field $E_{ex}$ in F, reduces the coherence length in
the ferromagnet, $\xi_{F}$, to $\hbar v_{F}/2E_{ex}$ in the clean
limit (where $v_{F}$ is the Fermi velocity) and to $\sqrt{\hbar D/2E_{ex}}$
in the dirty limit, both of which are typically of the order of 1 nm.

However, recent studies on NbTiN-$\textrm{\textrm{Cr\ensuremath{O_{2}}}}$-NbTiN
(SC-HMF-SC)\cite{Keizer2006} and Nb-$\textrm{C\ensuremath{u_{2}}MnAl}$-Nb
(SC-intermetallic F-SC) \cite{Sprungmann2010_trpl_JJ} Josephson junctions
revealed a long range PE where the supercurrent between the two SC
electrodes was measured even when the F-layer thickness was much larger
than $\xi_{F}$. Keizer et al.\cite{Keizer2006} attributed their
findings to the formation of a triplet pairing component at the $\textrm{Cr\ensuremath{O_{2}}}$
\textendash{} NbTiN interfaces, as predicted by Bergeret, Volkov and
Efetov\cite{BergeretVolfetovPRL86} and by Eschrig et al.\cite{EschrigPRL90}.
The long range Josephson effect reported by Sprungmann\cite{Sprungmann2010_trpl_JJ}
was similarly attributed by Linder and Sudbo\cite{LinderSudboJJ2010}
to triplet pairing, formed due to spin-active zones in the Nb-$\textrm{C\ensuremath{u_{2}}MnAl}$
interfaces. A triplet component with parallel spins can support AR,
in a similar manner as in singlet pairing in N-SC junctions, giving rise
to a PE on a length scale comparable to $\xi_{N}$. The orbital symmetry
of the induced triplet pairing correlations may be either even (s-wave,
d-wave) or odd (p-wave, f-wave), corresponding, respectively, to an
odd or even dependence on the Matsubara frequency\cite{BergeretVolfetovReview2005,EshcrigLofwanderHMF2008}.

Long range Josephson coupling was found also for junctions consisiting
of the high temperature superconductor $\textrm{YB\ensuremath{a_{2}}C\ensuremath{u_{3}O_{7-\delta}}}$
(YBCO) and the itinerant ferromagnet $\textrm{SrRu\ensuremath{O_{3}}}$(SRO)
\cite{AntognazzaAPL63}. Subsequently, STM measurements \cite{AsulinPRB74care}
suggested that this long range proximity effect (LRPE) may take place
only locally, along the domain walls (DWs) of the SRO. The tunneling
dI/dV \textit{vs}. V spectra revealed proximity superconducting gaps
on the surface of SRO layers much thicker than the expected $\xi_{F}$,
but these were confined to strips of width consistent with the DW
size of SRO. Thus, they attributed the local LRPE to the crossed Andreev
reflections (CARE) process\cite{ByersFlatteCARE1995,DeutscherCARE2000},
whereby a hole impinging on the interface in one magnetic domain is
retro-reflected as an electron with opposite spin polarization in
an adjacent domain having opposite magnetization. It should be noted
that this process can take place only if the DW width is up to a few
times larger than coherence length in the SC side of the junction,
$\xi_{S}$\cite{HerreraCAREat5Xi2009}, which is $\sim2$ nm in YBCO.
Volkov and Efetov\cite{VolfetovPEinF}, on the other hand, attributed
this LRPE to the formation of an odd-frequency triplet s-wave pairing
component in the DWs of F.

In order to check whether the CARE process is indeed essential for LRPE we studied
in this work bilayers of YBCO coated by the HMF $\textrm{L\ensuremath{a_{2/3}}C\ensuremath{a_{1/3}}Mn\ensuremath{O_{3}}}$
(LCMO). The DW width in thin LCMO films at 4.2 K is estimated\cite{LCMODWwidth2001}
to be $\sim20$ nm (although this value may depend on substrate, preparation
process and film thickness), which is much larger than $\xi_{S}$
in YBCO and thus the CARE process should be largely suppressed. Surprisingly,
our tunneling spectra revealed spectroscopic features conforming to
proximity-induced superconductivity even on LCMO layers as thick as
30 nm, an order of magnitude larger than the FFLO predicted $\xi_{F}$.
Moreover, these features were not confined only to narrow strips (as
in our previous study\cite{AsulinPRB74care} of SRO/YBCO bilayers)
and \textit{both} gaps and zero bias conductance peaks (ZBCP) were
measured. Our data suggest, as discussed below, that proximity-induced
triplet-pairing in LCMO plays an important role in the PE observed
in our samples. In a recent experiment, evidence for proximity induced
triplet pairing at the SC side of F/SC (In/Co or In/Ni) junctions
was provided \cite{AlmogTrplinSC2009}. In our present work, on the
other hand, the proximity induced triplet correlations are found on
the F side.

\section*{Experimental }

\begin{figure}
\includegraphics[width=0.5\textwidth]{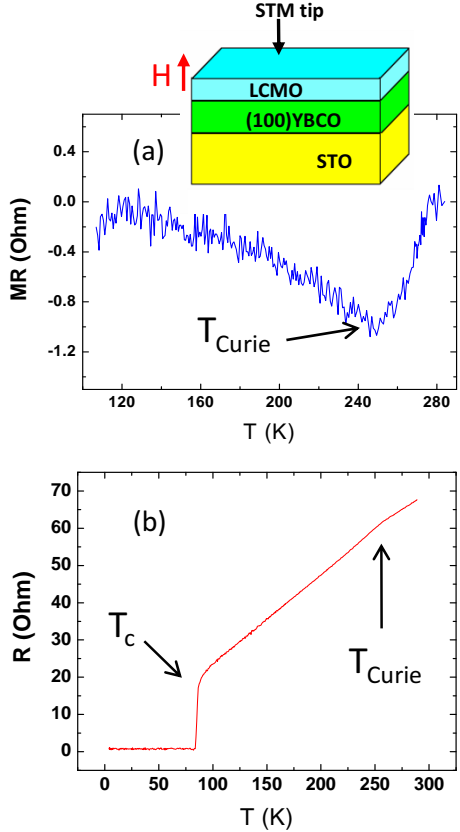}
\caption{\label{fig:RT}(color online). (a) Magneto-resistance R(2T)-R(0) measurement
of a 20 nm LCMO/(100)YBCO bilayer as a function of temperature showing
the ferromagnetic transition at $\sim250$ K. The inset shows the
structure of our samples, the direction of the magnetic field applied
in the magneto-resistance measurement and the tunneling direction in the STM measurements
(performed at H=0). (b) Resistance vs. temperature curve of the same
LCMO/YBCO bilayer depicting a superconducting transition at $\sim88$
K. The LCMO ferromagnetic transition is also manifested here, in the
slope change at $\sim250$ K. }
\end{figure}

We have studied epitaxial bilayers consisting of LCMO films deposited
on optimally doped \textit{a}-axis (100)YBCO films grown by laser
ablation deposition on $(100)\textrm{SrTi\ensuremath{O_{3}}}$ (STO)
substrates. The YBCO was deposited on the STO wafer in two steps.
First a 45 nm thick template layer of YBCO was deposited at a wafer
temperature of 600 $^\circ C$. Then a second 90 nm thick YBCO layer was prepared
at 750 $^\circ C$. The surface area of the YBCO consisted of 95\% (100)YBCO
(verified by x-ray diffraction) with no traces of (110) orientation
found. The LCMO layers were grown on top of the YBCO at 790 $^\circ C$ and
annealed for one hour at 450 $^\circ C$ in 0.5 Atm oxygen environment. The
whole process was performed \textit{in situ} without breaking the
vacuum, and given the good lattice matching between YBCO and LCMO,
high interface transparency could be obtained, important for the AR process to
take place. The thickness of our LCMO layers ranged between 15 nm
to 50 nm with surface roughness of $\sim1$ nm on top of the YBCO
crystallites. Full coverage of the YBCO by LCMO was confirmed by x-ray
photoelectron spectroscopy (XPS) and time of flight secondary ion
mass spectrometry measurements. Magneto-resistance and resistance
\textit{vs}. temperature measurements of the samples reveal a ferromagnetic
transition of the LCMO layer at $\sim250$ K and the SC transition
of YBCO at $\sim88$ K, as presented in Fig. \ref{fig:RT}. It is
important to note that \textit{a}-axis YBCO films were chosen for
this study, rather than the more common \textit{c}-axis (001)YBCO
films, since the PE is significantly suppressed along the \textit{c}-axis\cite{DurusoyKapitulnikPE1996,SharoniPEAuYBCO2004}.%

Samples were transferred in dry atmosphere to our cryogenic STM after
being exposed to ambient air for less than 10 minutes. After evacuation
the STM chamber was filled with He exchange gas at 1 Torr and then
cooled down to 4.2 K where all the measurements presented here were
performed, using a Pt-Ir tip. Several control measurements were performed
at temperatures up to 150 K to verify that the spectroscopic features
associated with SC (gaps and ZBCPs) indeed vanish above the $\textrm{\ensuremath{T_{c}}}$
of YBCO. Topographic images were taken in the standard constant current
mode with bias voltages $\sim100$ mV, well above the SC gap. The
tunneling dI/dV \textit{vs}. V spectra, which are proportional to
the local density of states (DOS), were numerically derived from the I-V
curves acquired on the LCMO surface while momentarily disconnecting
the STM feed-back loop.

\section*{Results}

\begin{figure}
\includegraphics[width=0.5\textwidth]{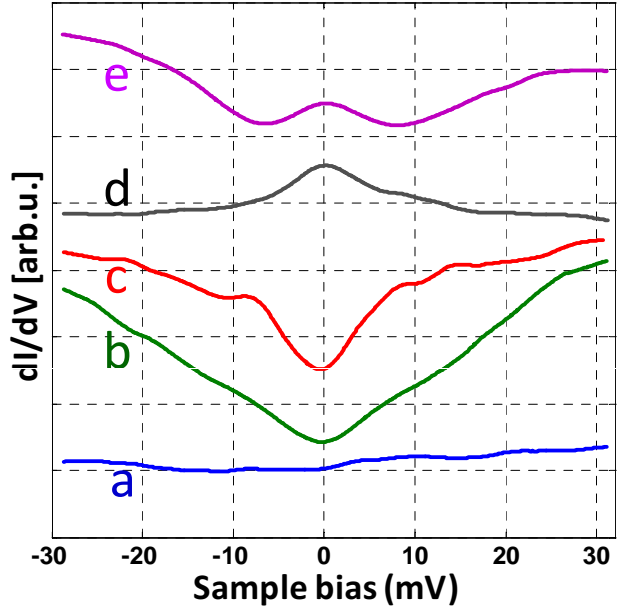}

\caption{\label{5spectra}(color online). Tunneling dI/dV vs. V spectra acquired
at 4.2 K on LCMO/(100)YBCO bilayers showing the five main different
spectral features measured on these F/SC bilayers (shifted for clarity).
Curves (a) and (b) show no signature of a PE, where presumably, curve (a) was
acquired over a metallic LCMO region and curve (b) was taken on a
more insulating region. The upper three curves manifest proximity-induced
SC order in the LCMO layer. Curve (c) shows the most prominent such
spectral feature that appeared in our measurements, a SC-like gap
in the DOS, whereas (d) and (e) portray ZBCPs, indicative of a sign-changing
(\textit{p}-wave or \textit{d}-wave) order parameter. The ZBCPs and
gaps appeared even on LCMO layers of 30 nm thickness (curve (d)) but
not on 50 nm thickness and not above $T_{c}$, indicating a long-range
proximity effect that we attribute to induced triplet pairing.}

\end{figure}

Tunneling dI/dV \textit{vs}. V curves representing the most typical
spectroscopic features found on our samples are portrayed in Fig.
\ref{5spectra}. Curves (a) and (b) do not show any SC-like features.
Curve (a) exhibits a nearly constant metallic-like DOS while curve
(b) exhibits a wide insulator-like gap structure. These spectra
are very similar to those reported in Ref \cite{ChenLCMOphases},
corresponding to conductive and insulating regions, respectively,
coexisting in LCMO films. Regions exhibiting such features were found
on all our samples, with abundance that grew with increasing LCMO
thickness. However, in many regions smaller gaps, having suppressed
SC coherence peaks (the so called \textquoteleft{}gap-like features\textquoteright{})
at 8-10 mV were observed (c), indicative of PE due to the SC YBCO film.
Interestingly, ZBCPs also appeared within such \textquoteleft{}gapped
areas\textquoteright{}, although less frequently. The ZBCP were found
to be either embedded within a gap-like feature (curve (e)) or not
(curve (f)), and in some cases showed splitting (see below). We recall
here in passing that the ZBCP is one of the hallmarks of anisotropic
sign-changing order parameters, such as \textit{p}-wave or \textit{d}-wave.
The SC-like spectroscopic features were observed on large areas of
the 15-20 nm thick LCMO samples, but their abundance decreased with
LCMO thickness. No such features were detected for the 50 nm LCMO thickness
bilayer, on which only the aforementioned metallic- and insulating-like
spectra were measured. It is also important to note that SC-like features
were not observed above the $\textrm{\ensuremath{T_{c}}}$ of YBCO
for all bilayers, indicating that they are indeed associated with
the PE.

\begin{figure}
\includegraphics[width=0.2\textwidth]{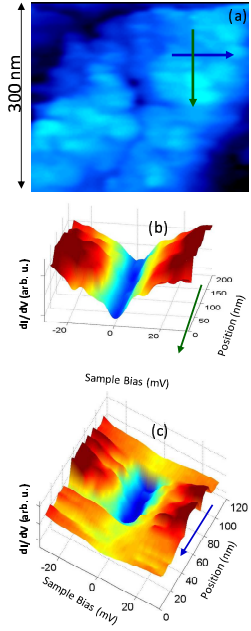}

\caption{\label{DW}(color online). STM measurement of the evolution of the
local DOS (dI/dV vs. V tunneling spectra) along two transverse lines
on a 15 nm LCMO/(100)YBCO bilayer. The gaps in the DOS along the green
and blue lines are shown in (b) and (c) respectively. These data suggest
that the LRPE in this region may be mediated by a domain wall. The
width of the 'gapped' region in (c), $\sim40$ nm, is of the order
of the domain wall width in LCMO, suggesting that in this case the
LRPE is mediated by the presence of a domain wall.}

\end{figure}

\begin{figure}
\includegraphics[width=0.3\textwidth]{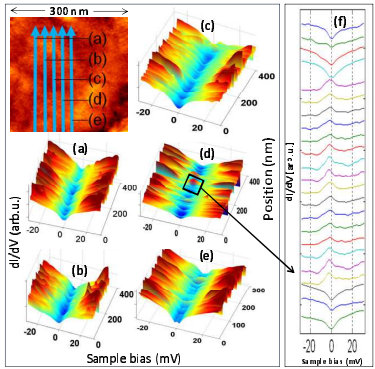}

\caption{\label{GapsAllOver}(color online). STM measurements of the local
DOS of a 17 nm LCMO/(100)YBCO bilayer showing SC-like features that
are not confined to domain walls. {[}(a)-(e){]} tunneling spectra
taken along the indicated (representative) lines.
Gaps in the DOS have been observed in areas of over 500nm x 100nm,
far wider than LCMO\textquoteright{}s typical domain wall width of
$\sim20$ nm. (f) Tunneling spectra (shifted for clarity) taken along the marked part of line (d) showing ZBCPs that appeared over a typical length scale of 10 nm.}

\end{figure}

We shall now discuss the spatial distribution of regions where SC-like
features appeared. On few occasions the gaps were confined to well-defined
strips, $\sim40$ nm wide, bordered by regions exhibiting metallic-like
spectra (Ohmic I-V curves), as shown in Fig. \ref{DW}. The resemblance
of the line width to that of the DW in LCMO suggests that in these
cases the spectra were acquired over a DW where the PE may be locally
enhanced, as found in Ref. \cite{AsulinPRB74care}. We note that the
locations of such lines showed no correlation with any specific topographic
feature. However, most commonly gaps were not confined to such narrow
lines and appeared to be spread over much larger areas, as demonstrated
by Fig. \ref{GapsAllOver}. Typically, these regions were surrounded
by areas characterized by the wide \textquoteleft{}insulating gap\textquoteright{}
(such as curve (b) in Fig. \ref{5spectra}), where the LCMO
was presumably more insulating and thus less likely to allow PE. The
spectra acquired along lines presented in Fig. \ref{GapsAllOver}
show that the proximity SC gaps varied spatially in both their width
and zero bias conductance. However, on average, the gaps became shallower
with increasing (nominal) LCMO layer thickness as depicted in Fig.
\ref{GapFilling}. Since the gaps\textquoteright{} zero bias conductance
varied spatially within each sample (see Figs. \ref{DW} and \ref{GapsAllOver}),
we present in Fig. \ref{GapFilling}, for each LCMO thickness, an
average spectrum calculated over regions of a few 100 $nm^{2}$ where
gaps having the lowest zero bias conductance (deepest gaps) were observed.

In contrast to the previous report by Asulin et al. \cite{AsulinPRB74care},
where only proximity gaps were observed for the SRO/YBCO bilayers,
in the present study also ZBCPs appeared within \textquoteleft{}gapped
regions\textquoteright{}, as demonstrated by Fig. \ref{GapsAllOver}(d
and f). No unique topographical feature that could be associated with
the nodal (110)YBCO surface was found in those regions. The ZBCPs
were significantly less abundant in our tunneling spectra compared
to the gaps, and moreover, they were not always robust and sometimes
disappeared upon repeating the I-V measurement at the same spot, turning
into gaps. This may be due to the effect of the measurement itself,
but nevertheless further indicates that the ZBCPsdid not arise due to any YBCO
faceting effect, and suggests that PE is the common origin of both
ZBCPs and gaps in the DOS measured on the LCMO surface.

In some cases more complex spectra were measured, such as very small
peaks within gaps, as shown in Fig. \ref{mixedOP}(a) and split ZBCPs,
with splitting of a few meV (Figs. \ref{mixedOP}(b) and \ref{mixedOP}(c)).
Interestingly, the spectra presented in each panel of Fig. \ref{mixedOP}
were taken along a single line of length less than 300 nm, yet large
variations from pure gaps or ZBCPs to the more complex structures
are observed. As will be further discussed below, the complex spectra
may indicate co-existence of two kinds of pairing symmetries, as predicted
in Ref. \cite{EshcrigLofwanderHMF2008}.

\section*{Discussion }

Our measurements clearly show that LRPE is a robust phenomenon existing
over large parts of LCMO/(100)YBCO bilayers. As discussed earlier,
a mechanism based on the FFLO model cannot account for PE to distances
larger than a few nm. In addition, the CARE-based LRPE scenario discussed
in Refs. \cite{AsulinPRB74care,KorenAronov2005,HerreraCAREat5Xi2009}
is also inapplicable here since the length scale over which the magnetization
direction can change, and in particular the DW width, is much larger
than the coherence length in YBCO. We note in this regard that the
CARE effect is predicted to be very small even when the DW width is
$5\xi_{S}$ ($\sim10$ nm in YBCO), quite narrower than the DW in
LCMO. Therefore, induced triplet-pairing in LCMO appears to be the
most probable mechanism by which PE is mediated in our samples.

\begin{figure}
\includegraphics[width=0.5\textwidth]{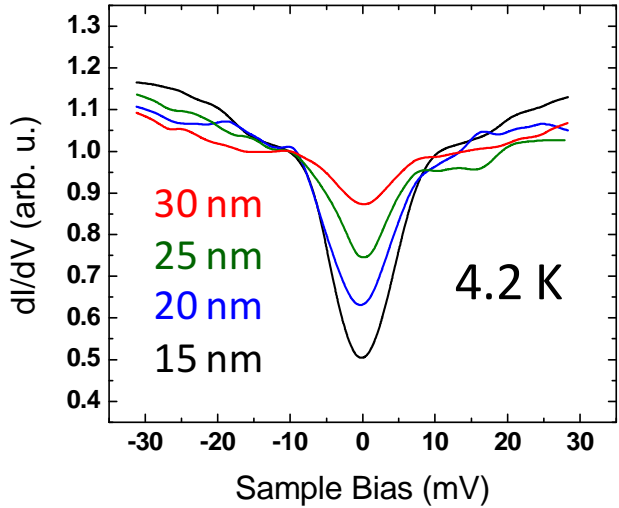}

\caption{\label{GapFilling}(color online). Averaged tunneling spectra at 4.2
K measured on a LCMO/YBCO bilayer with varying LCMO thickness, as
indicated. Each curve represents an average over spectra acquired
at regions where the gaps were most pronounced, having the lowest
zero bias conductance.}

\end{figure}

An interesting question that arises now is the orbital symmetry
of the induced triplet-pairing OP in the LCMO film. According to the
Pauli principle, if the Cooper pair wave-function is symmetric in
spin space, as in a triplet state, then it must be asymmetric in momentum
space. Consequently, triplet-pairing SCs are assumed to have \textit{p}-wave
orbital symmetry. Such a dependence on momentum, however, makes superconductivity
sensitive even to non-magnetic impurity (Anderson's theorem\cite{SchriefferBookSC})
and should not survive in disordered systems. Indeed, superconductivity
in ferromagnetic $\textrm{S\ensuremath{r_{2}}Ru\ensuremath{O_{4}}}$,
where singlet pairing is prohibited due to the exchange field, was
found to be of triplet \textit{p}-wave pairing, and has been observed
only in clean samples \cite{SigristSROpwave1998}. However, another
mechanism for triplet pairing (originally proposed by Berezinsky \cite{Berezinskii}
for superfluid $^{3}He$) enables the pair wave-function to abide
by Pauli's principle by being symmetric in momentum space and an odd
function of the Matsubara frequency. A corresponding odd frequency
spin-triplet \textit{s}-wave pairing is thus not sensitive to impurity
scattering and can therefore survive in F over a length scale of the
order $\xi_{N}\gg\xi_{F}$, like in \textquoteleft{}conventional\textquoteright{}
N-SC proximity systems. Bergeret, Efetov and Volkov showed that such
pairing can be promoted by magnetic inhomogeneity, either at the F/SC
interface\cite{BergeretVolfetovPRL86,BergeretVolfetovReview2005}
or at DWs \cite{VolfetovPEinF}, giving rise to LRPE. Eschrig et al.
\cite{EschrigPRL90,EshcrigLofwanderHMF2008} addressed the problem
of LRPE in HMF-SC junctions. They model a spin active interface causing
spin-mixing and breaking of the spin rotation symmetry. Such an interface
is formed, for instance, if there is a misalignment of the magnetic
moment at the F/SC interface with respect to that of the F bulk. In
this scenario, OPs of even frequency \textit{p}- or \textit{f}-wave
and odd frequency $s$- or \textit{d}-wave symmetries are induced in
the HMF with relative amplitudes that depend on the amount of disorder.

Proximity-induced odd-frequency triplet \textit{s}-wave pairing in
the LCMO side of our bilayers can well account for the gaps in the
DOS we observed on large regions of our samples having LCMO thicknesses
as large as 30 nm (see Fig. \ref{GapFilling}). According to Ref.
\cite{BergeretVolfetovPRL86} the SC correlations are expected to
penetrate the F up to a distance over which the magnetization changes its orientation which is of the order of the DW width, ($\sim20$
nm in LCMO), consistent with our findings. This mechanism\cite{VolfetovPEinF}
may also apply to the measurements in which gaps appeared along lines
conforming to the LCMO\textquoteright{}s DW structure. The variations
in the gap width and zero bias conductance along the surface, and
in particular the presence of areas showing metallic-like and insulating-like
DOS along with regions exhibiting clear proximity SC spectral features
may be due to spatial variations in the LCMO/YBCO interface transparency,
LCMO film properties, and the possible presence of an underlying \textit{c}-axis
YBCO crystallite (where PE is suppressed).

\begin{figure}
\includegraphics[width=0.5\textwidth]{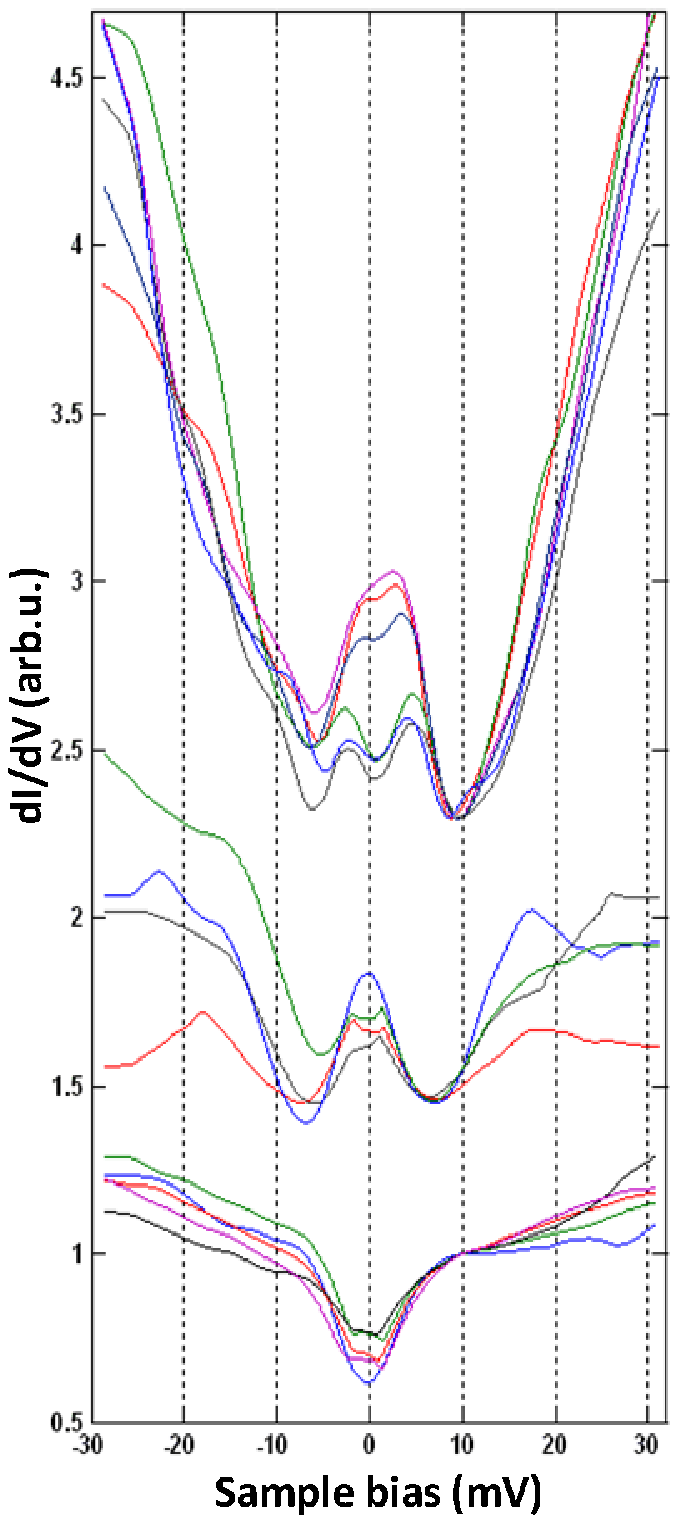}

\caption{\label{mixedOP}(color online). Tunneling spectra acquired along lines
of length smaller than 100 nm on 17 nm LCMO/YBCO bilayer (a) and 20 nm
LCMO\textbackslash{}YBCO bilayer {[}(b) and (c) - (shifted for clarity after normalization to the dip at positive voltage){]},
showing transitions from pure gap structure to a small peak within
a gap (a) and evolution of ZBCP splitting {[}(b) and (c){]}. These
data may be attributed to mixed induced-OP symmetries (\textit{s}-
\textit{p}- and \textit{d}-wave) with relative amplitudes varying along the sample.}

\end{figure}

Even more intriguing than the proximity induced gaps are the ZBCPs
that were found, although less abundantly compared to the gaps, on
all our bilayers with LCMO thickness up to 30 nm. Odd triplet \textit{s}-wave
OP cannot account by itself for the appearance of ZBCPs since an anisotropic
OP that changes its sign at the Fermi surface is required for their
formation (and the correction to the DOS at zero bias predicted in
Ref.\cite{VolfetovPEinF} is much smaller than our measured ZBCPs).
The two most probable candidates for such an OP have either \textit{d}-wave
or \textit{p}-wave symmetry, where ZBCPs are found for tunneling along
the nodal or anti-nodal directions, respectively\cite{Tanaka-d-wave1995,tanakahiwayaTunnelingTrplSC1997,WeiTunnelingPwave2005}. In the case of \textit{d}-wave symmetry it was shown\cite{PanZnImpurity2000} that ZBCPs may also arise in \textit{c}-axis tunneling due to impurity scattering.
As discussed above, the ZBCPs that we measured on LCMO are most probably not associated
with nodal (110) facets of the underlying \textit{d}-wave YBCO SC.
We thus attribute them to the penetration of either odd-frequency triplet
\textit{d}-wave or even-frequency triplet \textit{p}-wave order into the LCMO
film by either of the ZBCP-formation mechanisms discussed above. It is possible that the small and rather scarce regions where
ZBCPs appeared contained less disorder than other parts of the sample
and thus anisotropic pairing could survive. As mentioned earlier,
the ZBCPs were rather fragile, and they sometimes disappeared after
measurements at a specific location, turning into gaps. This suggests
that ZBCPs are indeed related to an unstable OP that may be affected
by the possible influence of the STM measurement on local disorder.
It should be noted here that although the current typically applied
in STM measurements is quite low, reaching about 0.1 nA in our case,
the corresponding current densities are rather high and therefore
STM measurements can become quite perturbative. Unfortunately, our
spectra do not allow us to state whether the ZBCPs are due to induced
\textit{d}-wave or \textit{p}-wave pairing, since there are many parameters
that one can play with in fitting the data to the extended BTK models
for tunneling into \textit{d}-wave\cite{Tanaka-d-wave1995} or \textit{p}-wave\cite{tanakahiwayaTunnelingTrplSC1997,WeiTunnelingPwave2005}
OP symmetries. Nevertheless, in some cases the rather large width
of the ZBCP conforms better to \textit{p}-wave pairing (such as the
black curve in Fig. \ref{5spectra}).

In addition to spectral features showing pure gaps or ZBCPs, indicative
of a single component OP, we have also measured more complex spectral
structures, such as peaks embedded within gaps or split ZBCPs, as
demonstrated by Fig. \ref{mixedOP}. Remarkably, the transition from
the 'simple' to 'complex' structures, as well as significant spatial
evolution of the latter, namely, development of the embedded peak
or ZBCP splitting, took place over distances of the order of 10-100
nm. The emergence of the ZBCP within the gap, shown in Fig. \ref{mixedOP}(a),
may be due to changing the tunneling direction with respect to the
main symmetry axes of the (\textit{d}-wave or \textit{p}-wave) OP.
This, however, cannot account for the ZBCP splitting. More likely,
the spectra presented in Fig. \ref{mixedOP} may reflect the formation
of a complex OP, namely, contributions with spatially varying degrees
of relative magnitude of \textit{s}-, \textit{p}- and \textit{d}-wave
components, possibly with phase shifts between them that may yield
ZBCP splitting\cite{FogelstromTunn_BRTS1997,Laughlin_dxy1998,TsueiPairSymmRev200,LinderSudboArxiv2010}.
We note in passing that spatial variations of the sub-dominant component
(\textit{s} or $d_{xy}$) of the complex OP \textit{d}+i\textit{s} or \textit{d}$\textrm{+i\ensuremath{d_{xy}}}$,
was recently observed by Ngai et al. on (110)$\textrm{Y\ensuremath{{}_{0.95}}C\ensuremath{a_{0.05}}B\ensuremath{a_{2}}C\ensuremath{u_{3}O_{7-\delta}}}$\cite{NgaiMixedOP2010}.
Alternatively, there have also been theoretical suggestions that the
spontaneous peak splitting can arise extrinsically, from either electron-hole
asymmetry, multiband effects or magnetic impurity perturbation \cite{GolubovAR,TanakashiwayaZBCPsplMltibnd2003,TanakashiwayaZBCPspltSctr2004}.
Further studies are needed to resolve these issues.

\section*{Summary}

In this paper we have used scanning tunneling spectroscopy to investigate
the local density of states on the LCMO surface of LCMO/(100)YBCO
(HMF/SC) bilayers. We found clear evidence for long range proximity
effect manifested as gaps and ZBCPs in the tunneling spectra. These
effects survived for LCMO film thicknesses of up to $\sim30$ nm, an
order of magnitude larger than the coherence length $\xi_{F}$ in LCMO
expected from the standard FFLO theory. Unlike the case of a previous
study on SRO/YBCO bilayers\cite{AsulinPRB74care}, the CARE mechanism
cannot account for the LRPE due to the large width of DWs in LCMO.
Triplet pairing in the HMF LCMO is suggested to be the underlying
mechanism for the observed LRPE. Gaps in the DOS abundant over very
large portions of our samples can be accounted for by odd frequency triplet
\textit{s}-wave pairing, which can survive over long distances due
to its insensitivity to impurity. The surprising appearance of ZBCPs
along with more complex SC-like features in some regions suggest
that the induced OP in LCMO has also \textit{p}-wave or \textit{d}-wave
components along with the \textit{s}-wave OP, with relative magnitude
determined by the local impurity concentration.

\section*{Ackowledgements }

This research was supported in parts by the joint German-Israeli DIP
Project, the United States\textemdash{}Israel Binational Science Foundation
Grant No. 200808, the Harry de Jur Chair in Applied Science, and the Karl Stoll Chair in advanced
materials.

\bibliographystyle{./apsrev4-1}
\bibliography{yoav_V3}

\end{document}